\documentclass[sigconf,nonacm]{acmart}
\usepackage{amsmath}

\setcopyright{cc}

\usepackage{etoolbox}
\usepackage{xurl}
\usepackage{amsmath}

\usepackage{wrapfig}

\usepackage{xspace}
\usepackage{xcolor}
\usepackage{graphicx}

\usepackage[inline]{enumitem}
\setlist{topsep=2pt}

\newtheorem{definition}{Definition}
\newtheorem{example}{Example}

\newcommand{\FACTS}{\ensuremath{F}\xspace}
\newcommand{\PROP}{\ensuremath{\mathrm{PROP}}\xspace}
\newcommand{\PLit}{\ensuremath{\mathrm{PLit}}\xspace}
\newcommand{\LIT}{\ensuremath{\mathrm{Lit}}\xspace}

\newcommand{\LABELS}{\ensuremath{\mathrm{Lab}}\xspace}

\newcommand{\Cost}{\ensuremath{\mathsf{C}}\xspace}
\newcommand{\Obl}{\ensuremath{\mathsf{O}}\xspace}
\newcommand{\Perm}{\ensuremath{\mathsf{P}}\xspace}

\newcommand{\TO}{\rightarrow}
\newcommand{\To}{\Rightarrow}
\newcommand{\defeater}{\leadsto}
\newcommand{\str}{\ensuremath{\mathrm{s}}\xspace}

\newcommand{\non}{\ensuremath{\mathcal{\sim }}}

\newcommand{\set}[2][\relax]{\ensuremath{#1\{#2#1\}}}
\newcommand{\Set}[2][\relax]{%
  \ensuremath{
    \ifx#1\left
      #1\{#2\right\}
    \else\ifx#1\right
      \left\{#2#1\}
      \else #1\{#2#1\}\fi\fi}%
}

\newcommand{\OBL}{\Obl}

\def\dRule#1:#2=>#3#4{#1\colon #2\To_{#3}#4}

\begin{document}

\title{Explainability by design: an experimental analysis of the legal coding process}

\author{Matteo Cristani}
\orcid{}
\email{matteo.cristani@univr.it}
\affiliation{
    \institution{University of Verona}
    \city{Verona}
    \country{Italy}
}

\author{Guido Governatori}
\orcid{0000-0002-9878-2762}
\affiliation{%
  \institution{Central Queensland University}
  \city{Rockhampton}
  \state{Queensland}
  \country{Australia}
}
\email{g.governatori@cqu.edu.au}
 
\author{Francesco Olivieri}
\affiliation{%
  \institution{}
  \city{Brisbane}
  \country{Australia}}
\email{}
\orcid{}

\author{Monica Palmirani}
\orcid{}
\email{monica.palmirani@unibo.it}
\affiliation{
    \institution{University of Bologna}
    \city{Bologna}
    \country{Italy}
}

\author{Gabriele Buriola}
\orcid{}
\email{gabriele.buriola@univr.it}
\affiliation{
    \institution{University of Verona}
    \city{Verona}
    \country{Italy}
}

\begin{abstract}
Behind a set of rules in Deontic Defeasible Logic, there is a mapping process of normative background fragments.
This process goes from text to rules and implicitly encompasses an explanation
of the coded fragments. 

In this paper we deliver a methodology for \textit{legal coding} that starts
with a fragment and goes onto a set of Deontic Defeasible Logic rules, involving a set of 
\textit{scenarios} to test the correctness of the coded fragments. The methodology is illustrated by the coding process of an example text. We then show the results of a series of experiments conducted with
humans encoding a variety of normative backgrounds and corresponding cases in which we have measured the efforts made in the coding process, as related to some measurable features. To process these examples, a recently developed
technology, Houdini, that allows reasoning in Deontic Defeasible Logic, has been employed.

Finally we provide a technique to forecast time required in
coding, that depends on factors such as knowledge of the legal domain, 
knowledge of the coding processes, length of the text, and a measure of
\textit{depth} that refers to the length of the paths of legal references.
\end{abstract}

\keywords{Legal coding; Explainable AI; Deontic Defeasible Logic}

\maketitle

\section{Introduction}
\label{sec:introduction}
Legal coding \cite{Godfrey2024153,Witt2024293,doi:10.1080/17579961.2022.2113670} is the process of translating the law into a formal language. This is actually complex and cumbersome for a number
of reasons, including the complexity of the legal language, the specialisation
of the reasoning process in the legal domain, and the dominance of logical constraints
towards each other that is implicit in the expression of legal bounds.
Many scholars \cite{Batsakis201831} have argued that a fruitful strategy to understand legal coding could be based on the \textit{RegTech} approach.
The foundation of this proposal is that whenever an intelligent system, based on
rules, makes a decision, we can trust the decision when we know the reasoning
system used for decision making and understand the rules as a correct representation
of the domain aimed to be coded. Thus, this approach relies on the following conceptual steps:

\begin{itemize}
\item Starting from the legal text, provide a \textbf{code} onto a \textit{standard} coding
system, accepted by a large community of practice in the domain;
\item \textbf{Verify} the correctness of the code by \textit{tests} that are
applications of the coded fragment into relevant \textit{scenarios}.
\end{itemize}

Therefore, the following tools are needed:

\begin{enumerate}
\item A logical apparatus consistent with human legal reasoning. In this paper,
we adopt a commonly employed formalism known as \textit{Defeasible Deontic
Logic} as defined in a number of studies \cite{Governatori20181,Governatori201246}. A framework-independent language, LegalRuleML \cite{lrml,Palmirani2011298,Athan20133, Athan2015151} that maps Deontic Defeasible Logic as well as other formalisms is employed to provide computer-processable source for the devised rules;
\item A technology that efficiently implements the aforementioned apparatus. For this task, we resort to Houdini, a technology for deontic defeasible reasoning implemented in Java \cite{Cristani2022,Cristani20234214,Cristani2024217};
\item A coding methodology. In this paper, a simplified methodology employed for the sole purposes of the experiment is proposed.
\item An instrumental experimental apparatus to guarantee the correct measure of
the coded fragments. We espouse the paradigm of empirical software engineering that has been conceived for general measures \cite{Moher198265}.
\end{enumerate}

 For the sake of easiness in the presentation we describe the translation process targeting only Deontic Defeasible Logic, and not LegalRuleML. The steps towards such a translation are provided in \cite{Palmirani201891}. 

The rest of the paper is organised as follows: we briefly define the framework of Deontic Defeasible Logic in Section \ref{sec:prel}, while Section \ref{sec:guide} introduces some guidelines recommended to be used during the translation process, which is described in its experimental investigation, main topic of this study; Section \ref{sec:example} describes an example of translation, showing how the translation process could be performed; in Section \ref{sec:experiment} we describe the experiment is its design and purpose; Section \ref{sec:results} discusses the results of the aforementioned experiment, and Section \ref{sec:concl} takes some brief conclusions sketching further investigations.

\section{Introduction to Deontic Defeasible Logic}
\label{sec:prel}

The logical apparatus we shall utilise for our investigation is the Standard Defeasible Logic (SDL) \cite{Antoniou2001255}. We start by defining the language.

Let $\PROP$ be the set of propositional atoms, then the set of literals $\LIT = \PROP \cup \{\neg p\, |\, p \in \PROP\}$. The \emph{complementary} of a literal $p$ is denoted by $\non p$: if $p$ is a positive literal $q$ then $\non p$ is $\neg q$, if $p$ is a negative literal $\neg q$ then $\non p$ is $q$. Literals are denoted by lower-case Roman letters. Let $\LABELS$ be a set of labels to represent names of rules, which will be denoted as lower-case Greek letters.

A defeasible theory $D$ is a tuple $(\FACTS, R, >)$, where $\FACTS$ is the set of facts (indisputable statements), $R$ is the rule set, and $>$ is a binary relation over $R$.

$R$ is partitioned into three distinct sets of rules, with different meanings to draw different ``types of conclusions''. \emph{Strict rules} are rules in the classical fashion: whenever the premises are the case, so is the conclusion. We then have \emph{defeasible rules} which represent the non-monotonic part (along which defeaters) of the logic: if the premises are the case then typically the conclusion holds as well unless we have contrary evidence that opposes and prevents us to draw such a conclusion. Lastly, we have \emph{defeaters}, which are special rules whose purpose is to prevent contrary evidence to be the case. It follows that in DL, through defeasible rules and defeaters,  we can represent in a natural way exceptions (and exceptions to exceptions, and so forth). 

We finally have the superiority relation $>$ a binary relation among couples of rules, that is the mechanism to solve conflicts. Given the two rules $\alpha$ and $\beta$, we have $(\alpha, \beta) \in >$ (or simply $\alpha > \beta$), in the scenario where both rules may fire (can be activated), $\alpha$'s conclusion will be preferred to $\beta$'s.


A rule $\alpha \in R$ has the form $\alpha\colon A(\alpha) \leadsto C(\alpha)$, where: (i) $\alpha \in \LABELS$ is the unique name of the rule, (ii) $A(\alpha) \subseteq \LIT$ is $\alpha$'s (set of) antecedents, (iii) $C(\alpha) = l \in \LIT$ is its conclusion, and (iv) $\leadsto\in \set{\TO, \To, \defeater}$ defines the type of rule, where: $\TO$ is for strict rules, $\To$ is for defeasible rules, and $\defeater$ is for defeaters.

Some standard abbreviations. The set of strict rules in $R$ is denoted by $R_s$, and the set of strict and defeasible rules by $R\str$; $R[l]$ denotes the set of all rules whose conclusion is $l$. 

A \emph{conclusion} of $D$ is a \emph{tagged literal} with one of the following forms:

\begin{description}
	\item[$\pm\Delta l$] means that $l$ is \emph{definitely proved} (resp. \emph{strictly refuted/non provable}) in $D$, i.e., there is a definite proof for $l$ in $D$ (resp. a definite proof does not exist).
	
	\item[$\pm\partial l$] means that $l$ is \emph{defeasibly proved} (resp. \emph{defeasibly refuted}) in $D$, i.e., there is a defeasible proof for $l$ in $D$ (resp. a definite proof does not exist).
\end{description}

The definition of proof is also the standard in DL. Given a defeasible theory $D$, a proof $P$ of length $n$ in $D$ is a finite sequence $P(1), P(2), \dots, P(n)$ of tagged formulas of the type $+\Delta l$, $-\Delta l$, $+\partial l$, $-\partial l$, where the proof conditions defined in the rest of this section hold. $P(1..n)$ denotes the first $n$ steps of $P$. 

For the sake of space we omit here the definitions of the proof tags and the corresponding proof conditions.

The last notions introduced in this section are those of extension of a defeasible theory. Informally, an extension is everything that is derived and disproved.

\begin{definition}[Theory Extension]
Given a defeasible theory $D$, we define the set of positive and negative conclusions of $D$ as its \emph{extension}:
\[
E(D) = (+\Delta, -\Delta, +\partial, -\partial ),
\]
where $\pm\# = \{l |\, l$ appears in $D$ and $D \vdash \pm\# l \}$, $\# \in \{\Delta, \partial\}$. 
\end{definition}

The deontic part of the logic encompasses obligations, permissions and prohibitions.
A prescriptive behaviour like ``At traffic lights it is forbidden to perform a U-turn unless there is a `U-turn Permitted' sign'' can be formalised via the general obligation rule
\[
\mathit{AtTrafficLight} \To_\Obl \neg \mathit{UTurn}
\]
and the exception through the permissive rule
\[
\mathit{UTurnSign} \To_\Perm \mathit{UTurn}.
\]

An alternative equivalent notation for permissions and obligations is to move the obligation or permission tag into the formula like below
\[
\mathit{AtTrafficLight} \To \Obl \neg \mathit{UTurn}
\]
and the exception through the permissive rule
\[
\mathit{UTurnSign} \To \Perm \mathit{UTurn}.
\]

The obligation of the negation of a literal establishes the prohibition of the opposite literal. Therefore the above determines that you are forbidden to pass on a red traffic light.

While \cite{GovernatoriORS13} discusses how to integrate strong and weak permission in 
Deontic Defeasible Logic, in this paper, we restrict our attention to the notion of strong permission, namely, when permissions are explicitly stated using permissive rules, i.e., rules whose conclusion is to be asserted as a permission.  

Following the ideas of \cite{ajl:ctd}, obligation rules gain more expressiveness with the \emph{compensation operator} $\otimes$ for obligation rules, which is to model reparative chains of obligations. Intuitively, $a \otimes b$ means that $a$ is the primary obligation, but if for some reason we fail to obtain, to comply with, $a$ (by either not being able to prove $a$, or by proving $\non a$) then $b$ becomes the new obligation in force. This operator is used to build chains of preferences (or repair chains), called $\otimes$-expressions.

The formation rules for $\otimes$-expressions are: (i) every plain literal is an $\otimes$-expression, (ii) if $A$ is an $\otimes$-expression and $b$ is a plain literal then $A \otimes b$ is an $\otimes$-expression \cite{GovernatoriORS13}. 

Summarising the proof conditions, we start with defining applicability and discardability, and following the structure of proofs for constituents and deontic formulas.

\begin{definition}[Applicability]\label{def:StandardApplicability}
	Assume a deontic defeasible theory $D = (\FACTS, R, >)$.  We say that rule $\alpha \in R^\Cost \cup R^\Perm$ is \emph{applicable} at $P(n+1)$, iff for all $a \in A(\alpha)$
	\begin{enumerate}
		\item\label{item:l} if $a \in \PLit$, then $+\partial_\Cost a \in P(1..n)$,
		\item if $a = \Box q$, then $+\partial_\Box q \in P(1..n)$, with $\Box \in\set{\Obl, \Perm}$,
		\item\label{item:Diamondl} if $a = \neg \Box q$, then $-\partial_\Box q \in P(1..n)$, with $\Box \in \set{\Obl, \Perm}$.
	\end{enumerate}

	We say that rule $\alpha \in R^\Obl$ is \emph{applicable at index} $i$ \emph{and} $P(n+1)$ iff Conditions~\ref{item:l}--\ref{item:Diamondl} above hold and
	\begin{enumerate}[resume]
		\item\label{item:cj} $\forall c_j \in C(\alpha),\, j < i$, then $+\partial_\Obl c_j \in P(1..n)$ and $+\partial_\Cost \non c_j \in P(1..n)$\footnote{\label{foot:violation}As discussed above, we are allowed to move to the next element of an $\otimes$-expression when the current element is violated. To have a violation, we need (i) the obligation to be in force, and (ii) that its content does not hold. $+\partial_\Obl c_i$ indicates that the obligation is in force.  For the second part we have two options. The former, $+\partial_\Cost\non c_i$ means that we have ``evidence'' that the opposite of the content of the obligation holds.  The  latter  would be to have $-\partial_\Cost c_j \in P(1..n)$ corresponding to the intuition that we failed to provide evidence that the obligation has been satisfied.  It is worth note that the former option implies the latter one. For a deeper discussion on the issue, see \cite{jurix2015burden}.}. 
	\end{enumerate}
	
\end{definition}

\begin{definition}[Discardability]\label{def:StandardDiscardability}
	Assume a deontic defeasible theory $D$, with $D = (\FACTS, R, >)$.  We say that rule $\alpha \in R^\Cost \cup R^\Perm$ is \emph{discarded} at $P(n+1)$, iff there exists $a \in A(\alpha)$ such that
	\begin{enumerate}
		\item\label{item:notl} if $a \in \PLit$, then $-\partial_\Cost l \in P(1..n)$, or
		\item if $a = \Box q$, then $-\partial_\Box q \in P(1..n)$, with $\Box \in \set{\Obl, \Perm}$, or
		\item\label{item:notDiamondl} if $a = \neg \Box q$, then $+\partial_\Box q \in P(1..n)$, with $\Box \in \set{\Obl, \Perm}$.
	\end{enumerate}

	We say that rule $\alpha \in R^\Obl$ is \emph{discarded at index } $i$ \emph{and} $P(n+1)$ iff either at least one of the Conditions~\ref{item:notl}--\ref{item:notDiamondl} above hold, or 
	\begin{enumerate}[resume]
		\item\label{item:notcj} $\exists c_j \in C(\alpha),\, j < i$ such that $-\partial_\Obl c_j \in P(1..n)$, or $-\partial_\Cost \non c_j \in P(1..n)$. 
	\end{enumerate}
	
\end{definition}
Note that discardability is obtained by applying the principle of \emph{strong negation} to the definition of applicability. The strong negation principle applies the function that simplifies a formula by moving all negations to an innermost position in the resulting formula, replacing the positive tags with the respective negative tags, and the other way around see \cite{DBLP:journals/igpl/GovernatoriPRS09}.
Positive proof tags ensure that there are effective decidable procedures to build proofs; the strong negation principle guarantees that the negative conditions provide a constructive and exhaustive method to verify that a derivation of the given conclusion is not possible. Accordingly, condition 3 of Definition~\ref{def:StandardApplicability} allows us to state that
$\neg\Box p$ holds when we have a (constructive) failure to prove $p$ with mode $\Box$ (for obligation or permission), thus it corresponds to a constructive version of negation as failure.  

We are finally ready to formalise the proof conditions, which are the standard in DDL \cite{GovernatoriORS13}. We start with positive proof conditions for constitutive statements. In the following, we shall omit the explanations for negative proof conditions, when trivial, reminding the reader that they are obtained through the application of the strong negation principle to the positive counterparts.

\begin{definition}[Constitutive Proof Conditions]\label{def:StandardCostProof}
\ 

\begin{tabbing}
  $+\partial_\Cost l$: \=If $P(n+1)=+\partial_\Cost l$ then\+\\
  (1) \= $l \in \FACTS$, or\\
  (2) \> (1) \= $\non l \not\in \FACTS$, and\\
      \> (2) \= $\exists \beta \in R^\Cost_\To[l]$ s.t. $\beta$ is applicable, and\\
      \> (3) \=$\forall \gamma\in R^\Cost[\non l]$ either\\
        \>\>(1) \=$\gamma$ is discarded, or \\
        \>\>(2)  $\exists \zeta \in R^\Cost[l]$ s.t. \\
            \>\>\> (1) $\zeta$ is applicable and \\
            \>\>\> (2) $\zeta > \gamma$.
\end{tabbing}  

\begin{tabbing}
  $-\partial_\Cost l$: \=If $P(n+1)=-\partial_\Cost l$ then\+\\
  (1) \= $l \not\in \FACTS$ and either\\
  (2) \= (1) \= $\non l \in \FACTS$, or\\
  \> (2) $\forall \beta \in R^\Cost_\To[l]$, either $\beta$ is discarded, or\\
  \> (3) \=$\exists \gamma\in R^\Cost[\non l]$ such that\+\+\\
        \=(1) $\gamma$ is applicable, and\\
        \=(2) $\forall \zeta \in R^\Cost[l]$, either $\zeta$ is discarded, or $\zeta \not > \gamma$.
\end{tabbing}
\end{definition}
A literal is defeasibly proved if: it is a fact, or there exists an applicable, defeasible rule supporting it (such a rule cannot be a defeater), and all opposite rules are either discarded or defeated.  To prove a conclusion, not all the work has to be done by a stand-alone (applicable) rule (the rule witnessing condition (2.2): all the applicable rules for the same conclusion (may) contribute to defeating applicable rules for the opposite conclusion. Note that both $\gamma$ as well as $\zeta$ may be defeaters.

\begin{example}\label{ex:Standard}
	Let $D = (F = \set{a, b, c, d, e}, R, {>}  = \set{(\alpha, \varphi), (\beta, \psi)})$ be a theory such that 
\begin{align*}
    R = \{&\alpha\colon a \To_\Cost l &&\beta\colon b \To_\Cost l && \gamma\colon c \To_\Cost l\\ 
    &\varphi\colon d \To_\Cost \neg l && \psi\colon e \To_\Cost \neg l && \chi\colon g \To_\Cost \neg l \}.	 
\end{align*}
\end{example}
Here, $D \vdash +\partial_\Cost f_i$, for each $f_i \in \FACTS$ and, by Condition (1) of $+\partial$. Therefore, all rules but $\chi$ (which is discarded) are applicable: $\chi$ is indeed discarded since no rule has $g$ as consequent nor is a fact. The team defeat supporting $l$ is made by $\alpha$, $\beta$ and $\gamma$, whereas the team defeat supporting $\neg l$ is made by $\varphi$ and $\psi$. Given that $\alpha$ defeats $\varphi$ and $\beta$ defeats $\psi$, then we conclude that $D\vdash +\partial_\Cost l$. Note that, despise being applicable, $\gamma$ does not  effectively contribute in proving $+\partial_\Cost l$, i.e. $D$ without $\gamma$ would still prove $+\partial_\Cost l$.

Suppose to change $D$ such that both $\alpha$ and $\beta$ are defeaters. Even if $\gamma$ defeats neither $\varphi$ nor $\psi$, $\gamma$ is now needed to prove $+\partial l$ as Condition (2.2) requires that at least one applicable rule must be a defeasible rule.  
Below we present the proof conditions for obligations.

\begin{definition}[Obligation Proof Conditions]\label{def:StandardOblProof}
\ 

\begin{tabbing}
  $+\partial_\Obl l$: \=If $P(n+1)=+\partial_\Obl l$ then\+\\
  	$\exists \beta \in R^\Obl_\To[l,i]$ s.t.\\
  	(1) \=$\beta$ is applicable at index $i$ and\\
  	(2) $\forall \gamma \in R^\Obl[\non l, j] \cup R^\Perm[\non l]$ either \+\\
  	   (1) \= $\gamma$ is discarded (at index $j$), or\\
  	   (2) \= $\exists \zeta \in R^\Obl[l, k]$ s.t.\+\\ 
  	   	\= (1) $\zeta$ is applicable at index $k$ and\\ 
  	   	\= (2) $\zeta >  \gamma$.
\end{tabbing} 

\begin{tabbing}
  $-\partial_\Obl l$: \=If $P(n+1)=-\partial_\Obl l$ then\+\\
  	$\forall \beta \in R_\To^\Obl[l,i]$ either\\
  	(1) $\beta$ is discarded at index $i$, or\\
  	(2) \= $\exists \gamma \in R^\Obl[\non l, j] \cup R^\Perm[\non l]$ s.t. \+\\
  	   (1) \= $\gamma$ is applicable (at index $j$), and\\
  	   (2) \= $\forall \zeta \in R^\Obl[l, k]$ either\+\\ 
  	   	\= (1) $\zeta$ is discarded at index $k$, or\\
  	   	\= (2) $\zeta \not >  \gamma$.
\end{tabbing} 

\end{definition}
Note that: (i) in Condition (2) $\gamma$ can be a permission rule as explicit, opposite permissions represent exceptions to obligations, whereas $\zeta$ (Condition 2.2) must be an obligation rule as a permission rule cannot reinstate an obligation, and that (ii) $l$ may appear at different positions (indices $i, j,$ and $k$) within the three $\otimes$-chains.  The example below supports the intuition behind the restriction to obligation rules in Conditions (2.2).
 
\section{Coding guidelines}
\label{sec:guide}
The experimental pipeline is described in Section \ref{sec:experiment} and precisely in Figure \ref{fig:structure}. Generally speaking, let us describe the framework of our investigation. Since we aim at valuing precisely the time span required to execute a coding, the experimental apparatus is made to measure the quality of the translation (namely the performance indicators typical of prediction systems - essentially measuring the errors produced in the translation) and the time required to execute a coding by hands. To conduct this activity we organised a preliminary workshop on the topic, to instruct the involved people, and then provide the data and execute the coding. Further on, we tested the coding of both normative background and scenarios by means of an automated tool. The result is a measure that forecasts with good performance the time required to execute a coding.

Details on the organisation of the workshop are provided, along with the organisation of the experiments, in Section \ref{sec:experiment}. The encoding and testing methodology we introduce here is essentially formed by three phases:

\begin{enumerate}
\item The \textit{normative background encoding phase} in which a coder, starting from a legal text, generates a coded version of the law. This operation is here conducted towards Deontic Defeasible Logic, but in practice that would be much easier if implemented in LegalRuleML, for which there already exist editors with annotators. The presentation is limited for the sake of readability. The implemented rules shall be of four types: 
\begin{itemize}
\item \textit{Strict rules}, namely definitions, that are named, in the terminology of LegalRuleML, \textit{constituent rules}.
\item \textit{Propositional defeasible rules} aiming at capturing concrete aspects of the behaviour of the reality, such as physics, biology but also very elementary social rules, without obligations and permissions.
\item \textit{Superiorities} namely preferences expressed between two propositional or deontic defeasible rules to express the fact that when both the involved rules are activable (namely they have the antecedents all proven), the one that is actually activated is the preferred one.
\item \textit{Deontic Defeasible rules} that represent obligations, permissions and prohibitions.
\end{itemize}
\item The \textit{scenario encoding phase} in which a coder takes some examples of scenarios actually fitting a specific configuration of the legal background, and map them into the LegalRuleML language (in Deontic Defeasible Logic). In this context we shall have two types of coding of the scenario, one for the case of Private Law and one for Criminal. In the first case we need to match meta-rules \cite{Olivieri2024261, Malerba2022149,Rotolo2024116,Governatori2021181}, whilst in the second we employ flat Deontic Defeasible Logic, and we need to introduce numeric constraints and assignments for the construction of repair chains that include serve time for imprisonment or money for fines. This matter has been discussed in \cite{Cristani2022}.
\item The \textit{evaluation phase} in which the coder takes the coded examples and tests them against the normative background. The idea of the methodology is that the coder should compare the decision she would make about the case with the decision made by a machine, in particular the reasoner on which both the background and the scenario are inserted. When the decisions are not equal, we should mark this as a problem, that needs to be evaluated again. 
\end{enumerate}

At the end of the application of the methodology we have encoded, tested and evaluated a normative background and the result is also marked as pass/review depending on possible errors determined by scenarios with errors.

Obviously the methodology has some critical aspects that we need to propose in the very beginning:

\begin{itemize}
\item \textit{Differences in language}. Let us assume that we would like to encode a large mass of norms. It will be definitely impractical to commit one single person in doing so, therefore a reasonable solution to such an encoding problem will be to employ numerous members of an encoding team. However, this would provoke potential discrepancies in the process, for different coder would choose different terms to summarise the elements appearing in the domain. This risk could also be present in the coding of one single person, who does not detect the need to employ one single term for different synonyms or the negated term when an antonym appears. To mitigate this risk we should consider three tools: (1) a previously established glossary, with associated thesaurus and corresponding archetypical terms for synonyms and antonyms  (2) a set of guidelines, as defined below, to restrict the variances, and (3) a cyclic methodology in which the steps defined above are repeated until we reach a number of errors in the scenarios below a given threshold we consider acceptable for the context.
\item \textit{Implicitness of defeasible rules}. It could be the case that a coder considers obvious that the development of a specific deontic rule lies on a known condition. For instance, an assertion such as \textit{it is forbidden to run in an area denominated quiet} does not apply to people who are not able to run, but if you implicitly assume that someone on a wheelchair does not run, and do not put this condition on a rule, the exclusion is not computed. To mitigate this we should employ some longer time on encoding from empirical observations, for instance by choosing this role for specific members of the encoding team.
\end{itemize}

The recommendation we provided during the instruction workshop can be summarised in the following points: 

\begin{enumerate}[label=\alph*)]
\item When encountering a noun phrase any adjective attached to the noun that would naturally form a specialisation of the noun itself should be taken, along with the  noun, as a single positive literal. For instance, when translating \textit{divorced spouse}, we devise \textit{divorced$\_$spouse} as the actual literal.
\item When a copulative verb (for instance \textit{to be} or \textit{to become}) appears in the sentence, the subject is predicated onto the nominal part in form of a definition by means of strict rules. For instance, the sentence \textit{A sale is a contract} is translated into \textit{sale $\rightarrow$ contract}.
\item When a verb different from copulative ones occurs in the sentence it has to be attached to subject and complements to form a single literal. For instance \textit{Mario buys Arsenic} translates into \textit{Mario\_buys\_Arsenic}
\item When encountering a modal of obligation or a modal of permission they have to remain attached onto the noun or the noun phrase as it has been generated. Analogously, when a punishment is provided as a consequence of a specific behaviour, we shall introduce a \textit{repair chain} that is introducing the punishment as a literal. When a punishment literal is introduced the repair literal is the head of a propositional rule assigning a value to a variable accounting for the punishment duration. Moreover, when generic expressions are used to devise a general obligation or permission, we recommend to use a deontic rule without tail. For example \textit{Whoever causes the death of a person is punished with no less than twenty-one years} is translated as in the basic example of Section \ref{sec:example}.
\item Any conditional sentence, even implicitly expressed by verbs such as causes, generates and others of this very same nature is to be translated into an defeasible rule, possibly with deontic labels, and potentially repair chain if punishements are introduced as well.

\end{enumerate}

The above guidelines do not constitute automation rules (in fact, the experimental activities in automated translation for legal coding have been quite disappointing so far). As we shall see, in practice, despite the guidelines, individual variances are still present.

\section{An example of translation}
\label{sec:example}

We introduce here two coding examples. The first example starts with a legal text excerpted from the Italian Criminal Law. The text has been used as source in the translation. We provide the full text requested to be translated by the experimenters in the collection of documents  which can be made available on request. We show the translation process only for the basic article of the normative background, and for the exception that applies here. This text (in the Italian version) is the number 1 of Table \ref{tab:ground} \\ \ \\
\fbox{\begin{minipage}{8cm}{
\begin{center} \textit{Art. 575 (Italian Criminal Code) - Homicide}\end{center}
Whoever causes the death of a person is punished with imprisonment of no less than twenty-one years.}\end{minipage}}
\\ \ \\ \noindent
Exceptions - Aggravating Circumstances:\\
The penalty of \textbf{life imprisonment} applies when the crime is committed:
\begin{itemize}

\item Using poisonous substances or other insidious means;
\item With premeditation.
\end{itemize}

The translation process generates a Deontic Defeasible Logic expression, that is generated by the experimenters, while directed to adhere to the above specified guidelines.

Article 575 can be translated as follows (the reported one is the actual translation by one of the experimenters). \\ \ \\
\fbox{
\begin{minipage}{8cm}
      $\Rightarrow \OBL \sim$ death $\otimes$ basic$\_$punishment\\
    basic$\_$punishement $\Rightarrow$ imprisonment := 21years
\end{minipage}
}
\\ \ \\

For the exceptions, we adopt the principle that the translation should be on two steps. Firstly we map the deontic rule as it would have been a primary one, and not an exception. In the above case we shall have the translations below.\\ \noindent

\fbox{
\begin{minipage}{8cm}
    poisonous\_means  $\Rightarrow \OBL \sim$ death $\otimes$ life$\_$imprisonment\\
    premeditation  $\Rightarrow \OBL \sim$ death $\otimes$ life$\_$imprisonment\\
\end{minipage}
}
\\ \ \\

On top of the normative background, as we devised above, we provided also cases, namely descriptions of circumstances in which it would be possibly applied one of the above mentioned codes. In this case we describe a single case corresponding to homicide in order to introduce the exact procedure.
\\

\textit{The two brothers, Alberto and Mario, have been living together since their parents' death and have never married nor had stable relationships. Despite living together for a long time, they have never found a good balance, and in fact, Alberto mistreats Mario, patronizing him, forcing him to do household chores, and often mocking him for his clumsiness.
After many years of this harsh life, Alberto falls ill, loses the use of his legs, and ends up in a wheelchair. From that moment on, the mistreatment increases, accompanied by fits of uncontrollable rage, shouting, screaming, and insults. After a few weeks, Mario reaches a breaking point, and unable to endure his brother any longer, he buys a bottle of arsenic, mixes it into Alberto's food, and within three days, causes his death.}

On request of judgment by human experimenters, this situation is translated into the applicability of the article 575 and the aggravating circumstances of usage of poisonous means and premeditation. In practice, the translation is a set of facts and propositional rules: \\ \ \\ 
\noindent
\fbox{
\begin{minipage}{8cm}
Alberto\\
Mario\\
living$\_$together\\
Alberto$\_$mistreats$\_$Mario\\
ill$\_$Alberto\\
wheelchair$\_$Alberto\\
rage$\_$Alberto$\_$Mario\\
Mario$\_$buys$\_$Arsenic\\
Mario$\_$poisons$\_$Alberto\\
Alberto, Mario, Mario$\_$buys$\_$Arsenic, \\Mario$\_$poisons$\_$Alberto $\Rightarrow$ Death\\
Alberto, Mario, Mario$\_$buys$\_$Arsenic, \\Mario$\_$poisons$\_$Alberto $\Rightarrow$ Poisonous$\_$means\\
Alberto, Mario, Mario$\_$buys$\_$Arsenic, \\Mario$\_$poisons$\_$Alberto $\Rightarrow$ Premeditation
\end{minipage}
}
\\ \ \\
Due to the facts and the rules we can compute a set of propositional consequences, the \textit{extension} of this propositional theory, and while applying the deontic rules coding the homicide and the aggravating circumstances we derive, logically, that Mario shall be punished with life imprisonment.

To clarify about variances, we note that in this specific case the experimenters reached contradictory conclusions for this case. One of them, in fact, argued that the mistreats, the rage, shouts and all the patronizing and the harassments of Alberto against Mario, perpetrated for a very long time are sufficient to establish that there is a \textit{generic} (namely valid for any crime) mitigating circumstance, the named \textit{provocation}. This is likely, in the opinion of this experimenter, to be equivalent to the aggravating one of poisoning, but not likely to overwhelm the premeditation.

On private law, conversely, we shall have devise several definitions, much more than direct deontic rules with punishments, and rules employed to code, for instance, contracts. Specifically, one of the norms we required to code is below, and it is taken from the Italian Private Law Code.
\\ \ \\ \noindent
\fbox{\begin{minipage}{8cm}{
\begin{center} \textit{Art. 1470 Italian private law code}
\end{center}The sale is the contract that aims to transfer the ownership of a thing or the transfer of another right in exchange for the payment of a price.

}\end{minipage}}

\ \\ \noindent Again we have the case.

\textit{
Celeste decided to purchase a set of knives after seeing them in a TV infomercial. Marvellous Blade, the company selling them, claimed they were the sharpest on the market, capable of cutting through cobblestones and drainpipes without sustaining any damage. Sceptical yet intrigued, Celeste called their number and proposed to buy the knives only after personally testing their extraordinary qualities. Confident in their bold claims—and likely motivated by a shortage of customers—the Marvellous Blade representative agreed to her conditions and sold her the knife set.
}

\ \\ \ \noindent The totality of experimenters classified the verbal agreement reached by Celeste and Marvellous Blade a sale.

\section{An experiment in legal coding}
\label{sec:experiment}
The example of coding process described in Sec. \ref{sec:example} has been conducted with the express purpose of providing a code for a text, with the corresponding explanations (namely translations in Deontic Defeasible Logic). The translation team was chosen on purpose without complete expertise in the matter (namely nobody was an both an expert in Deontic Defeasible Logic and in legal matters), in order to look at translations neutrally. At a purely qualitative level, we can observe that the search for an explanation is a relevant issue in the process, but mainly, it drives to a coding process that can be then observed on its actual performance.

\begin{figure*}{htb!}
\includegraphics[width=0.9\linewidth]{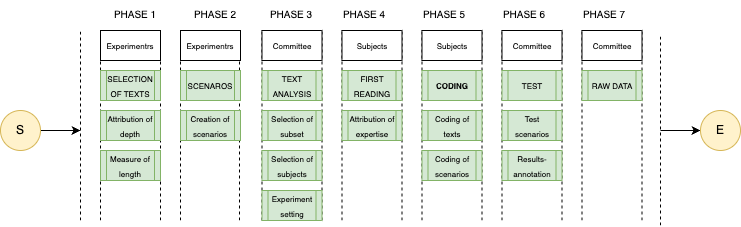}
\caption{\label{fig:structure} The experimental design.}
\end{figure*}

This coding activity, in other terms, is an observable behaviour of a human coding activity, and therefore, it can be measured with some software engineering metrics.
We should observe, however, that though we may start with a number of subjects and deliver some parameters of measure in front (time to code is the most relevant), more difficult is to obtain some form of correlation with parameters that regard the humans employed in the process. To investigate the aforementioned aspects, we devised the following experiment whose structure is reported in Figure \ref{fig:structure}.

\subsection*{Preparing the setting}

We arranged eleven legal texts taken from both the penal and the private codes of the Italian normative background. Each of these texts was then adjoined with the relevant related documents, in particular those that constitute legal references. All the above operations have been conducted by three expert jurists (the experimenters), in order to isolate in the best possible way the relevant variables to be measured. The jurists were required to address the texts with an explicit information on the \textit{depth}, that it is measured on a scale of 5, where a text has depth 1 when it is part of the constitution or related implementation norms, depth 2 when it depends on the constitution and related norms only, and further on. We have investigated on a preliminary basis, about the maximum depth, with the aforementioned jurists who agreed on the value of five for the incumbent norms. Conventionally, texts that form a body by international treatises or conventions (e.g. the universal declaration of human rights) are assumed to have depth 0. Constitutional norms should be referred to depth 1, and depth 2 is reserved to general norms such as fundamental codes (civil, criminal, procedural). General norms are typically of depth 3, whilst norms of depth 4 are decrees and regulations, and finally depth 5 is reserved for techical annexes to the regulations.

Further, we selected seven out of the eleven texts, to balance length on a growing scale and guarantee simultaneously a good distribution of the samples on the depths. The results of this selection process are described in Table \ref{tab:ground}.

\begin{table}[h]
\begin{tabular}{|l|l|l|}
\hline
\textbf{Text} & \textbf{Length} & \textbf{Depth}\\ \hline
1	& 673	& 2\\ \hline
2	& 891	& 3\\ \hline
3	&1287	& 3\\ \hline
4	& 1455	& 2\\ \hline
5	& 1822	& 4\\ \hline
6	& 2011	& 3\\ \hline
7	&2344	& 3\\ \hline
\end{tabular}
\caption{\label{tab:ground} text lengths and depths for the experiment.}
\end{table}

After the selection process, we ask the experimenters to augment the texts with
six scenarios in which the normative background corresponding to the legal text result relevant, and identify the legal consequences of the background on those scenarios. The scenarios were chosen with a similar selection method from a larger corpus of 17.

\subsection*{The legal coding}

We selected a group for the experiments to perform legal coding. The selection group was formed by twenty three jurists and seven non-jurists. We organised a workshop during which the subjects were taught the bases of Deontic Defeasible Logic, in order to let them able to code correctly. After this phase we asked them to produce an experimental code, to value the ability and to code and the required time at a very gross level in order to plan the activities. Some of these experimenters were performing quite nicely in terms of coding of the normative background and in terms of coding of the scenarios. However, not each of these thirty ones were adequate.

Fourteen experimental subjects, among expert and less expert jurists and other individuals (in particular non-jurists with general knowledge of the coding process), were selected at this stage, who completed coding correctly for all the above. When a subject performed correctly on the experimental text, we asked them to propose a coding for a scenario. Those who passed this test where enrolled for second phase.

Subjects have been firstly proposed to read the texts and express a judgement on their expertise on the topics on a scale of percentage on decimals. Since the majority of jurists in Italy are oriented towards either penal or private backgrounds, we have chosen a bend of these, and some of these jurists have valued themselves, on request, quite differently on the topics related to the single texts. This could have resulted in a complexity in separating the variables to measure the performances; therefore, we have selected some non-jurists among the experts on coding. The resulting self-judgements are reported in Table \ref{tab:ground2}.

\begin{table}[h]
\begin{tabular}{|l|l|l|l|l|l|l|l|}
\hline
 \textbf{Subject} & \textbf{T. 1}& \textbf{T. 2}& \textbf{T. 3}& \textbf{T. 4}& \textbf{T. 5}& \textbf{T. 6}& \textbf{T. 7}\\ \hline
1	& 0.7	& 0.6	& 0.9	& 1.0	& 0.5	& 0.8	& 0.8 \\ \hline
2	& 0.7	& 0.9	& 0.9	& 1.0	& 0.5	& 0.8	& 0.8 \\ \hline
3	& 0.8	& 0.6	& 0.6	& 1.0	& 0.6	& 0.8	& 0.8 \\ \hline
4	& 0.8	& 0.5	& 0.6	& 1.0	& 0.6	& 0.9	& 0.9 \\ \hline
5	& 0.6	& 0.8	& 0.9	& 0.6	& 0.6	& 0.8	& 0.8 \\ \hline
6	& 0.6	& 0.9	& 0.9	& 0.6	& 0.6	& 0.8	& 0.8 \\ \hline
7	& 0.8	& 1.0	& 0.5	& 0.9	& 0.5	& 0.7	& 0.9 \\ \hline
8	& 0.7	& 1.0	& 0.9	& 0.9	& 0.6	& 0.8	& 1.0 \\ \hline
9	& 0.3	& 0.3	& 0.3	& 0.3	& 0.3	& 0.3	& 0.3 \\ \hline
10	& 0.2	& 0.2	& 0.2	& 0.2	& 0.2	& 0.2	& 0.2 \\ \hline
11	& 0.1	& 0.1	& 0.1	& 0.1	& 0.1	& 0.1	& 0.1 \\ \hline
12	& 0.1	& 0.1	& 0.1	& 0.1	& 0.1	& 0.1	& 0.1 \\ \hline
13	& 0.1	& 0.1	& 0.1	& 0.1	& 0.1	& 0.1	& 0.1 \\ \hline
14	& 0.1	& 0.1	& 0.1	& 0.1	& 0.1	& 0.1	& 0.1 \\ \hline
\end{tabular}
\caption{\label{tab:ground2} Experimental subjects and their expertise.}
\end{table}

We then settle the actual experiment, where the fourteen subjects above have been requested to code, on an individual base, all the seven chosen texts. They were given the opportunity to consult the normative background, as well as the rules' structure in Deontic Defeasible Logic, and their coding activity has been measured in terms of resulting length of the coded Deontic Defeasible Logic and the required time for the coding. After the coding process, a verification phase took place in which, exchanging the coded norms among jurists and non-jurists, subjects were allowed to listen to corrections and then use the consultancies to correct the results. The time required for the exchange and for the re-coding, estimated in an essentially standard time , has been summed up and reported along with the time for the individual coding. Overall, this phase has contributed in an essentially constant time and the reported recoding has been very low. We decided not to separate these phases in the accounting of performance time, since the values are too tight to constitute a significant measure. Besides, it is rather obvious that some consultancy and consequently recoding could occur in a large number of cases.

After the coding process, the subjects were asked to code the scenarios. In particular, we shuffled the subjects making the evaluation by choosing the subjects at random, with no subject to be chosen to code scenarios for her coding of the text, and subjects who make the coding to be able to access both the original text and the coded one. 

This has been a request that we misjudged in the preliminary setting of the experiment, wrongly assuming that the required time to code the scenarios would have been essentially irrelevant to the process. This was not the case. Coding the scenarios requires time which depends on the complexity of the legal background the scenarios refer to. We therefore resettle the experiment to appropriately measure the coding time for the scenarios. Essentially, coding the scenarios requires the same amount of time required for the encoding of the background per character.

\subsection*{Tests on the Scenarios}

This testing phase was conducted by a second group of experimenters, consisting of five people experts of the technological setup. Having the coded texts at disposal, the group performed the tests on the scenarios as coded by the subjects. This phase, in terms of time, has been totally irrelevant. The subjects provided their coded texts in a file format that can be directly uploaded to the Houdini system \cite{Cristani2024217,Cristani20234214,Pasetto2023,Cristani2022}. Some small mistakes have been made by the subjects and therefore the second group of experimenters needed some time to correct these errors, but the overall added time is not relevant to the experimental purpose. Moreover, this extra time is expected to be nullified by subsequent improvements in the interface of Houdini. In any case, the time can be estimated to be a 20$\%$ excess on the total time of code for background and scenarios.

Finally, we collected the correctness of the scenario analysis associating the number of erroneously coded scenarios to each norm coding. After this phase, we gave the results to the subjects and asked them to correct the errors in the scenarios. The large majority of these requests have been solved essentially in real time. The overall required time to be added is specifically proportional to the time required for the coding. In particular, the error rate is on average 6$\%$ and the required recoding on this phase is 5.8$\%$. This is neatly biased by the procedure whose limits are numerous.

\section{The results of the experiment: a general analysis}
\label{sec:results}
The result of the experiment can be summarised in the following concept: legal coding is a reliable process. In fact, the error rate is rather insignificant, with corrections that are straightforward and not time consuming as compared to the coding process itself.

Regarding length, the coded texts in formal language (Deontic Defeasible Logic) are on average 19$\%$ longer than the original texts (in number of characters). There is anyway, a significant standard deviation of 9$\%$ and a span from 2$\%$ to 33$\%$. However, while eliminating the two outliers of the extreme values the result is 20$\%$ with a standard deviation of only 3$\%$. This leads to the claim that the sample is too small to take a reasonable conclusion on the actual distribution and therefore we should derive a preliminary acceptance of the value per se. 

For what concerns the outcome in terms of performance, the results are more complex to analyse; we can start from a very general analytical point, namely the temporal analysis structure. We have measured time of coding while knowing the length of the texts started with, using as unique parameter the \textit{number of characters}. This choice is determined by the fact that this is an objective measure not depending on the chosen text format, or other more structural aspects (number of words, number of paragraphs or even articles). This analysis has given two synthesis indices that could be interesting to devise: mean and median. The result is that the time required to code spans from a minimum of 1.8  to a maximum of 5.91 seconds per character, with an average time of 3.99 seconds; the standard deviation is 0.94 seconds. On the other hand, the median is 4.06, very close to the average value. Applying the Shapiro-Wilk test on the population \cite{Shapiro19681343}, we get that the departure from normality is below 0.01 on a 0.05 confidence interval span. 

A part from being a very important validation of the experimental setting, for we can presuppose this to be verified a priori, the normal distribution is the best possible guess on a continuous variable and we have also that the distribution results in a relatively low span, with a low percentage on the average value. This means that we can consider the average value a good predictor of time required to code, with some corrections we shall discuss further on.

Since the average page contains 1800 characters, we can presuppose that the time required to code a page is roughly two hours (spanning from a minimum of 54 minutes to a maximum of 3h57').

As discussed in Section \ref{sec:experiment}, having collected data regarding expertise and depth we can now look at these values, aiming to understand whether there is an influence of the expertise in the delivery time. 

\begin{figure*}[htb!]
\includegraphics[width=0.8\linewidth]{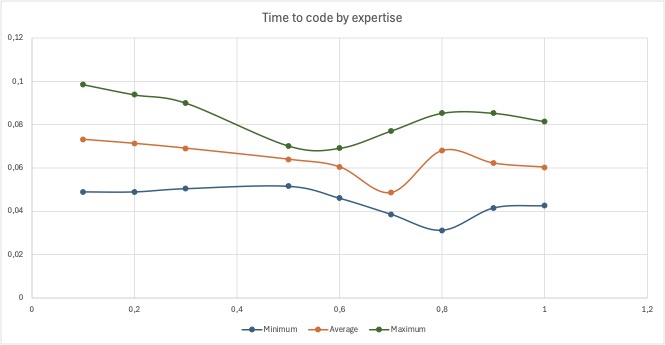}
\caption{\label{fig:exp}Time to deliver coding by classes of expertise.}
\end{figure*}

If we measure the correlation index between the time per character with the self-valued expertise, then we have a value of -28.67$\%$ compatible with the hypothesis of a weak inverse linear correlation with the expertise; namely experts code slightly quicker than non-experts. With a class partition based on decimals, we can also figure this out as reported in Figure \ref{fig:exp}, where the time of coding per character on average by classes of expertise from 0 to 1 is exposed.

A qualitative analysis of the graphical representation by classes gives two hints:

\begin{itemize}
\item Up to the expertise of 0.8 there is a rather evident negative correlation of the time required to code, that improves moderately while growing the expertise;
\item Once we pass the 0.8, the time worsen on scale, but still preserving a negative correlation on that portion of the domain. We may observe that, as usually happens for self-evaluations, there is a known tendency to self-esteem that is higher in lower expert (also well known by the name of Dunning-Kruger effect). This can have driven the most experts to under-evaluate themselves and produced the effect of this behaviour here.
\end{itemize}

Regarding, on the other hand, the relation between depth and time to code, there is again a very similar negative correlation, to -28.71$\%$ which in turn explains itself by the higher difficulty of a text with higher depth. 

Regarding bias analysis, we have to consider the following aspects:

\begin{enumerate}
\item The aforementioned point on the \textit{size} of the sample. We should mention that a post-coding longitudinal investigation based on years of actual coding with usage in practice could give a stronger value to a study like this, but it is nonsense to imagine an investigation that could require hundreds of subjects and thousands of hours of observation to guarantee the elimination of this bias to an acceptable confidence interval. The likelihood of the investigation lies on the method and we can claim we have a point here.

\item The point on the self-estimation method for both expertise and depth. These are quite challenging points to be addressed in order to eliminate the bias to a reasonable extent. Our view is that it is not worth getting onto such a complex path, for the ability of evaluating the expertise is so subjective that it would probably be required a number of double blind access evaluations that would hinder any possibility of success.

\item The point on language and specific normative background. This is unavoidable. It is almost impossible to devise a disposition of a test, even with the simple purpose of finding a coefficient to measure time to code for a text, that results independent on the normative background, and therefore the language. However, the quality of the adopted approach ensures that it could be repeated to compute other coefficients for different backgrounds and different languages as well. 
\end{enumerate}

A further analysis regards the variability factors of the translation operations. Although we provided, during the preparatory workshop mentioned above, a general methodology with some guidelines as described in Section \ref{sec:guide}, there are factors inevitably generating variability:

\begin{enumerate}

\item There are numerous, and difficult to classify a priori, ways to express modalities in natural language. These encompass the express usage of modal verbs, the usage of modal expressions (is entitled to, has the duty of, is responsible for). These variants can be perceived differently, depending on the context.

\item When a text is extended with related material, as in the first step of the methodology described above.

\item Some cases may arise in which multiple adjectives are attributed to a noun and this could lead to a different interpretation: either we mean to attribute both adjectives to the same noun to compound onto a single noun phrase or we do so while compounding separately in two distinct noun phrases.

\end{enumerate}

We should also report that the error rate is rather low, if compared to other experimental settings in the same context. This is certainly a bias on phases 6 and 7, since the same subjects coding the norms actually coded the scenarios to provide synchronization. Obviously this generates a bias. If different subjects would make a test on scenarios they code whilst the normative background is coded by others (even if the synchronization is provided a priori) the  error rate would go up. On the other hand alignment could be framed onto specific guidelines, and we could expect a long-term convergence process that in the end shall guarantee the performances.

\subsection{A scenario with the Italian Criminal Code}
If we employ the \textbf{basic} number extracted from the experiments, namely the number 4 that is the number of seconds to encode one single character, we can estimate the \textit{coding effort} for such an activity. Obviously this is not enough to determine the actual \textit{elapsed time} that need to be employed. This is more complicated to establish, but, based on the observations we made above, could be downmarked here.

Let us start with some raw data that shall help us in both the activities defined above. In Table \ref{tab:icc} we report raw data on the Italian Criminal Code.

\begin{table}[h!]
\begin{tabular}{|p{2.4cm}|p{3cm}|}
\hline
 Length &  435,939 characters\\ \hline
 Number of sections (libri) &  3\\ \hline
 Number of subsections (titoli) &  8, 14, 4\\ \hline
 Number of articles & 832 \\ \hline
 Typical depth &  1, 2\\ \hline
 Interference with other codes & Criminal procedure code, Italian Constitution  \\ \hline
 \end{tabular}
\caption{\label{tab:icc} Raw data on the Italian Criminal Code.}
\end{table}

By applying the basics of the methodology we can estimate the total effort in 485 hours. This is only the effort of encoding the background. A rough estimate, obtained by expert opinions, collected during the workshop, generates a length of the scenarios of likely at least twice the background. There exists a large collection of examples of these cases that can be considered (for instance the UK National Archives have a Section dedicated to this topic - https://www.nationalarchives.gov.uk/help-with-your-research/research-guides/criminal-court-cases-an-overview/), so the effort would consist in \textit{finding} right cases, that could be estimated in a further 100$\%$ of the total. Overall, we can go for the following estimate:

\begin{itemize}
\item Time to code: 485 hours;
\item Time to retrieve scenarios: 485 hours;
\item Time to code scenarios: 485*3=1455 hours;
\item Time to test: (485+1455)*0,2=388 hours.
\end{itemize}

The total time required is 2813 hours.
Assuming the coding and testing process is assigned to one single person, the total time to deliver coding and testing elapses to roughly 23 person months. The highest parallelization we can imagine is by subsections. The total number is 26, that are not all equivalent in terms of length, but for a first figure we can assume that this is the case and that will provide a much shorter effort, minimum getting into one single month. However, the parallelization would be definitely not possible in this way for three different reasons:

\begin{itemize}
\item We need to forecast time to sort discrepancies. This shall account to a very long excess, that is very difficult to determine a priori, without further data;
\item We need to understand better the internal interference in order to split the job. Possibly this is not the best split, for it generates a load in other phases that is difficult to forecast;
\item The testing phase will result more complicated if conducted by separate teams in complex cases.
\end{itemize}

To be noted is also that the large part of criminal case descriptions that can be found do not summarize the situation in a schematic way, and therefore not only time to code would result longer but also the test would result hard to forecast. Moreover, a limited number of these cases are complex for they correspond to more than one criminal behaviour in the same context, possibly involving more than one subject. This will generate a longer and more complex coding of the case and consequently a longer definition of the scenario and a longer test.

\section{Conclusions and further developments}
\label{sec:concl}

The experiment has given several important hints onto any future process of translation of the law based on scientific ground and solid engineering methodologies.

First of all, it is clear that we have by no means a technique that can be used to predict the required coding time for \textit{single pieces of the law}. The ability to measure coding time relies on \textit{large amounts of texts} and \textit{numerous translators}, since the individual variance is very significant; and the mean time per character, computed as a result of the investigation, only makes sense on a large base.

Secondly, the large variance referred above is mainly individual, depending on the expertise of the translator more than on the depth (in some sense, a measure of the complexity of references) of the text.

There is a flat ability of a translation methodology to work appropriately on the scenario that lies completely on the \textit{correspondence of the used translation tokens}, namely on the rigid application of some sort of schemata. This suggests that an important aspect to reduce variance would be to guarantee this integrity of language in the translation, for instance by giving as a means a standard setup of certain language tokens on which some consensus can be found among jurists.

We are in the process of integrating the results of this paper with a further study, where segregation methods (in particular unsupervised clustering) are employed to subdivide a legal text corpus into separated subcorpora where every single subcorpus can be coded by a subset of the encoding team, with a low risk of interference with the activities performed in the other subcorpora. This will enhance the possibility of finding some consensus onto the adopted language token set and reduce therefore individual variance, leaving the translation effort onto a large number of individuals, and thus making it feasible.

\bibliographystyle{plain}
\newcommand{\doi}[1]{\url{https://doi.org/#1}}

\bibliography{biblio}

\begin{thebibliography}{10}

\bibitem{doi:10.1080/17579961.2022.2113670}
Alice~Witt Anna~Huggins, Mark~Burdon and Nicolas Suzor.
\newblock Digitising legislation: connecting regulatory mind-sets and constitutional values.
\newblock {\em Law, Innovation and Technology}, 14(2):325--354, 2022.

\bibitem{Antoniou2001255}
G.~Antoniou, D.~Billington, G.~Governatori, and M.J. Maher.
\newblock Representation results for defeasible logic.
\newblock {\em ACM Transactions on Computational Logic}, 2(2):255--287, 2001.

\bibitem{Athan20133}
Tara Athan, Harold Boley, Guido Governatori, Monica Palmirani, Adrian Paschke, and Adam Wyner.
\newblock {OASIS} {LegalRuleML}.
\newblock In {\em Proceedings of the International Conference on Artificial Intelligence and Law}, pages 3--12, 2013.

\bibitem{Athan2015151}
Tara Athan, Guido Governatori, Adrian Paschke, Monica Palmirani, and Adam Wyner.
\newblock {LegalRuleML}: Design principles and foundations.
\newblock In Wolfgang Faber and Adrian Paschke, editors, {\em Reasoning Web. Web Logic Rules}, number 9203 in LNCS, pages 151--188. Springer, 2015.

\bibitem{Batsakis201831}
Sotiris Batsakis, George Baryannis, Guido Governatori, Ilias Tachmazidis, and Grigoris Antoniou.
\newblock Legal representation and reasoning in practice: A critical comparison.
\newblock In {\em Jurix 2018}, volume 313 of {\em Frontiers in Artificial Intelligence and Applications}, pages 31--40, 2018.

\bibitem{Cristani2022}
M.~Cristani, G.~Governatori, F.~Olivieri, L.~Pasetto, F.~Tubini, C.~Veronese, A.~Villa, and E.~Zorzi.
\newblock Houdini (unchained): an effective reasoner for defeasible logic.
\newblock In {\em Proceedings of 6th Workshop on Advances in Argumentation in Artificial Intelligence}, volume 3354 of {\em CEUR Workshop Proceedings}, 2022.

\bibitem{Cristani20234214}
M.~Cristani, G.~Governatori, F.~Olivieri, L.~Pasetto, F.~Tubini, C.~Veronese, A.~Villa, and E.~Zorzi.
\newblock The architecture of a reasoning system for defeasible deontic logic.
\newblock In {\em Procedia Computer Science}, volume 225, pages 4214--4224, 2023.

\bibitem{Cristani2024217}
M.~Cristani, F.~Olivieri, G.~Governatori, and G.~Buriola.
\newblock Simulating the law in a multi-agent system.
\newblock In {\em Proceedings of the 25th Workshop From Objects to Agents}, volume 3735 of {\em CEUR Workshop Proceedings}, pages 217--232, 2024.

\bibitem{Godfrey2024153}
Nicholas Godfrey and Mark Burdon.
\newblock Fidelity in legal coding: applying legal translation frameworks to address interpretive challenges.
\newblock {\em Information and Communications Technology Law}, 33(2):153 – 176, 2024.

\bibitem{jurix2015burden}
Guido Governatori.
\newblock Burden of compliance and burden of violations.
\newblock In Antonino Rotolo, editor, {\em 28th Annual Conference on Legal Knowledge and Information Systems}, Frontieres in Artificial Intelligence and Applications, pages 31--40, Amsterdam, 2015. IOS Press.

\bibitem{Governatori20181}
Guido Governatori.
\newblock Practical normative reasoning with defeasible deontic logic.
\newblock In Claudia d'Amato and Martin Theobald, editors, {\em Reasoning Web 2018}, number 11078 in LNCS, pages 1--25. Springer International Publishing, Cham, 2018.

\bibitem{Governatori2021181}
Guido Governatori, Francesco Olivieri, Antonino Rotolo, Abdul Sattar, and Matteo Cristani.
\newblock Computing private international law.
\newblock {\em Frontiers in Artificial Intelligence and Applications}, 2021.

\bibitem{GovernatoriORS13}
Guido Governatori, Francesco Olivieri, Antonino Rotolo, and Simone Scannapieco.
\newblock Computing strong and weak permissions in defeasible logic.
\newblock {\em J. Philos. Log.}, 42(6):799--829, 2013.

\bibitem{DBLP:journals/igpl/GovernatoriPRS09}
Guido Governatori, Vineet Padmanabhan, Antonino Rotolo, and Abdul Sattar.
\newblock A defeasible logic for modelling policy-based intentions and motivational attitudes.
\newblock {\em Logic Journal of the {IGPL}}, 17(3):227--265, 2009.

\bibitem{ajl:ctd}
Guido Governatori and Antonino Rotolo.
\newblock Logic of violations: A gentzen system for reasoning with contrary-to-duty obligations.
\newblock {\em Australasian Journal of Logic}, 4:193--215, 2006.

\bibitem{Governatori201246}
Guido Governatori, Antonino Rotolo, and Erica Calardo.
\newblock Possible world semantics for defeasible deontic logic.
\newblock In Thomas Ågotnes, Jan Broersen, and Dag Elgesem, editors, {\em 11th International Conference on Deontic Logic in Computer Science}, volume 7393 of {\em Lecture Notes in Computer Science}, pages 46--60, Heidelberg, 2012. Springer.

\bibitem{Malerba2022149}
Alessandra Malerba, Antonino Rotolo, and Guido Governatori.
\newblock A logic for the interpretation of private international law.
\newblock In Shahid Rahman, Matthias Armgardt, and Hans~Christian Nordtveit~Kvernenes, editors, {\em New Developments in Legal Reasoning and Logic}, volume~23 of {\em Logic, Argumentation and Reasoning}, pages 149--169. Springer International Publishing, Cham, 2022.

\bibitem{Moher198265}
Thomas Moher and G.~Michael Schneider.
\newblock Methodology and experimental research in software engineering.
\newblock {\em International Journal of Man-Machine Studies}, 16(1):65--87, 1982.

\bibitem{Olivieri2024261}
Francesco Olivieri, Guido Governatori, Matteo Cristani, Antonino Rotolo, and Abdul Sattar.
\newblock Deontic meta-rules.
\newblock {\em Journal of Logic and Computation}, 34(2):261--14, 2024.

\bibitem{lrml}
Monica Palmirani, Guido Governatori, Tara Athan, Harold Boley, Adrian Paschke, and Adam Wyner.
\newblock {LegalRuleML} core specification version 1.0.
\newblock Oasis committee specification, OASIS, 2020.

\bibitem{Palmirani2011298}
Monica Palmirani, Guido Governatori, Antonino Rotolo, Said Tabet, Harold Boley, and Adrian Paschke.
\newblock {LegalRuleML}: {XML}-based rules and norms.
\newblock volume 7018 of {\em Lecture Notes in Computer Science}, pages 298--312, Heidelberg, 2011. Springer.

\bibitem{Palmirani201891}
Monica Palmirani, Michele Martoni, Arianna Rossi, Cesare Bartolini, and Livio Robaldo.
\newblock Legal ontology for modelling gdpr concepts and norms.
\newblock In {\em Jurix 2018}, volume 313 of {\em Frontiers in Artificial Intelligence and Applications}, pages 91--100, 2018.

\bibitem{Pasetto2023}
L.~Pasetto, M.~Cristani, G.~Governatori, F.~Olivieri, and E.~Zorzi.
\newblock Extraction of defeasible proofs as explanations.
\newblock In {\em 7th Workshop on Advances in Argumentation in Artificial Intelligence}, volume 3546, 2023.

\bibitem{Rotolo2024116}
Antonino Rotolo and Giovanni Sartor.
\newblock Logical models for private international law.
\newblock In Michael~S. Green, Ralf Michaels, and Roxana Banu, editors, {\em Philosophical Foundations of Private International Law}. Oxford University Press, 2024.

\bibitem{Shapiro19681343}
S.S. Shapiro, M.B. Wilk, and H.J. Chen.
\newblock A comparative study of various tests for normality.
\newblock {\em Journal of the American Statistical Association}, 63(324):1343--1372, 1968.

\bibitem{Witt2024293}
A.~Witt, A.~Huggins, G.~Governatori, and J.~Buckley.
\newblock Encoding legislation: a methodology for enhancing technical validation, legal alignment and interdisciplinarity.
\newblock {\em Artificial Intelligence and Law}, 32(2):293--324, 2024.

\end{thebibliography}

\end{document}